\documentclass[useAMS,usenatbib,fleqn]{mn2e}

\usepackage{amsmath}
\usepackage{graphicx}
\usepackage{subfigure}
\usepackage{bm}

\newcommand{\apj}{ApJ}
\newcommand{\aj}{AJ}
\newcommand{\icarus}{Icarus}
\newcommand{\aap}{A\&A}
\newcommand{\araa}{ARAA}
\newcommand{\mnras}{MNRAS}
\newcommand{\apjl}{ApJL}
\newcommand{\jspin}{j_\mathrm{s}}
\newcommand{\jorb}{j_\mathrm{o}}
\newcommand{\ciso}{c_\mathrm{iso}}
\newcommand{\spin}{\Omega_\mathrm{s}}
\newcommand{\brk}{\Omega_\mathrm{b}}
\newcommand{\ssize}{$\langle r_*\rangle$ }

\title[Star-disc interactions]{Spin down of protostars through
  gravitational torques}  
\author[Lin, Krumholz \& Kratter]{Min-Kai Lin,$^{1,2}$ 
\thanks{mkl23@cam.ac.uk} 
Mark R. Krumholz$^{1}$\thanks{krumholz@ucolick.org} and Kaitlin
M. Kratter$^{3}$\thanks{kkratter@cfa.harvard.edu} \\  
$^{1}$Department of Astronomy,
University of California, Santa Cruz, CA 95064, USA\\
$^{2}$Department of Applied Mathematics and Theoretical Physics,
University of Cambridge, Cambridge CB3 0WA, U.K\\
$^{3}$Institute for Theory and Computation, Harvard-Smithsonian Center
for Astrophysics, Cambridge, MA 02138, USA}

\begin{document}
\maketitle

\begin{abstract}
Young protostars embedded in circumstellar discs accrete from an
angular momentum-rich mass reservoir. Without some braking mechanism,
all stars should be spinning at or near break-up velocity.  In this
paper, we perform simulations  of the self-gravitational collapse
of an isothermal cloud using the ORION adaptive mesh refinement code
and investigate the role that gravitational torques might play in the
spin-down of the dense central object. While magnetic effects
likely dominate for low mass stars, high mass and Population III stars
might be less well magnetised. We find that gravitational torques
alone prevent the central object from spinning up to more than
half of its breakup velocity, because higher rotation rates lead to
bar-like deformations that enable efficient angular momentum transfer
to the surrounding medium. We also find that the long-term spin
evolution of the central object is dictated by the properties of
the surrounding disc. In particular, spiral modes with azimuthal
wavenumber $m=2$ couple more effectively to its spin than the lopsided
$m=1$ mode, which was found to inhibit spin evolution. We suggest that
even in the absence of magnetic fields, gravitational torques may 
provide an upper limit on stellar spin, and that moderately massive  
circumstellar discs can cause long-term spin down.
\end{abstract}


\section{Introduction}\label{intro}
One of the unsolved problems in the physics of star formation is the
spin of protostars \citep{bodenheimer95}. The observed specific angular
momentum of molecular clouds typically exceeds that of stars by as
much as four orders of magnitude \citep{goodman93}. Most of the modern
attempts at explaining the removal of excess angular momentum from the
protostar invoke magnetic torques or magnetic stellar winds
\citep[e.g.][and references therein]{matt05, matt08, matt10}.
Although some theoretical studies on the effect of purely hydrodynamic
star-disc interactions on stellar spin have been carried out
\citep{yuan85,popham91,kogan93,glatzel99}, non-magnetic  mechanisms
have largely been neglected. However, there may exist situations where
magnetic fields are unavailable to remove angular momentum.

Population III stars have long been thought to be unmagnetised due to 
both the absence of seed fields, and the high temperatures required to 
maintain a sufficient ionisation fraction in  metal free gas
\citep{tan04}. However, these ideas have recently been 
challenged \citep{federrath11,schleicher11}, and the
issue remains unresolved. Recent simulations suggest that without the
action of magnetic fields or gravitational torques, the first stars
might reach near break-up velocities \citep{stacy11}.

Even for present-day massive star formation it is unclear that the
magnetic mechanisms normally invoked to regulate stellar spins are
applicable. Massive stars form with much higher accretion rates than
low-mass ones, and their discs tend to be dominated by gravitational
rather than magnetic angular momentum transport mechanisms 
  \citep{krumholz07,kratter08,krumholz09,hennebelle11,peters11}. 
Naively inserting the high accretion rates typical of massive star
formation  into the models most commonly adopted to explain T Tauri
star spin rates \citep[e.g.][]{matt08} yields the conclusion that
massive stars should be spinning at break-up \citep[][in
preparation]{rosen11}, in contrast with observed stellar spins
\citep{wolff06}. Recent observations of discs around young B stars are
also consistent with non-magnetised accretion columns, in contrast to
their lower mass counterparts \citep{eisner10}. Thus,
  non-magnetic braking mechanisms are also of interest for
  limiting stellar spin. Since the most general shape a protostar
  can have is triaxial, gravitational torques from a protostar acting
  on its surroundings should be explored.

 Gravitational
  torques as an important means of angular momentum transport have been
  proposed to solve the angular momentum problem of cloud cores in an
  analytic work by \cite{fisher04} and a numerical study by
  \cite{jappsen04}. However, these authors consider multiple stellar
  systems. Instead, the simpler problem we consider --- that of a single
  object interacting with its surroundings --- is closer to the setup
  envisioned by  \cite{yuan85}. Using existing theories, they
  discussed how the spin angular momentum of a protostar may
  be regulated by the circumstellar disc. If the protostar rotates
  rapidly, it may be subject to instability and become triaxial. It
  will then exert a torque on the disc, allowing the protostar to transfer
  angular momentum to the disc. To the best of our knowledge, no numerical
  simulations similar to \citeauthor{yuan85}'s problem have yet been
  undertaken.  However, transport of angular momentum via gravitational torques has been
observed in simulations focusing on the first core phase of star formation
\citep{bate98, saigo06, saigo08, saigo11}.

In this work, we present calculations of the collapse of an
  isothermal sphere and study the role of  gravitational torques
acting on the central object to remove  its spin angular
momentum. By modelling the central object as a finite-volume
fluid body, we demonstrate through simple numerical experiments that
its spin can be limited by its deformation and the gravitational
interaction with the surrounding medium. In addition, we find that
long term spin-down is possible if the surrounding disc develops
$m=2$ spiral modes, but spin-evolution  is inhibited if the disc
develops significant $m=1$ non-axisymmetry.

This paper is organised as follows. In \S\ref{model} we describe our
model setup and numerical method. We define diagnostic measures in
\S\ref{diagnostics}. We compare and contrast two simulations in  
\S\ref{results}, pointing out important correlations
between disc modes and spin-down. We discuss several important caveats
to our conclusions in \S\ref{caveats}. In 
\S\ref{conclusions} we discuss implications of our results on
protostellar spins and stellar evolution.

\section{Star-disc model} \label{model}
We consider the collapse of a cloud leading to the formation of a
dense central object surrounded by a disc. For convenience we will
  use `star' and `stellar' to refer to the central object and its
  properties, even though the problem as 
  we set it up is fully dimensionless, so the central object does not
  necessarily correspond to the physical size or internal structure of a
  real star. We discuss this issue further in \S\ref{dimensionless}.

The system has mass $M_\mathrm{sys} = M_d + M_*$, where $M_d$
and $M_*$ are the disc and stellar masses respectively. A recent
numerical study \citep{kratter10} shows that discs formed
via this idealised collapse are characterised by two dimensionless
parameters describing its mass accretion and rotation rates.    

Accretion of material onto the disc from the cloud is described by the
parameter $\xi$: 
\begin{align}
  \xi \equiv \frac{G\dot{M}}{\ciso^3},
\end{align}
where $\dot{M}$ is the mass accretion rate and $\ciso$ is the
isothermal sound-speed. The cloud rotation, responsible for disc
formation, is described by the parameter $\Gamma$: 
\begin{align}
  \Gamma \equiv \frac{\dot{M}}{M_\mathrm{sys}\Omega_k},
\end{align}
where $\Omega_k$ is the Keplerian orbital frequency of material
joining the system from the cloud, assumed to occur at cylindrical
radius $R_k$ such that $\Omega_k^2 = GM_\mathrm{sys}/R_k^3$. It can be
shown that \begin{align}\label{kepler_radius} 
  R_k = \xi^{1/3}\Gamma^{2/3}\ciso t= h^2\xi \ciso t,
\end{align}
where $M_\mathrm{sys} = \dot{M}t$ has been used, $t$ is the elapsed
time and $h$ is the disc aspect-ratio at $R_k$. In this work we
specify $h$ instead of $\Gamma$ directly.  

The system is modelled as a single, non-magnetic, inviscid and
self-gravitating fluid with density $\rho$, pressure $P$, velocity
$\bm{v}$ and gravitational potential $\Phi$. Its evolution is governed
by the usual Euler equations and the Poisson  equation. To distinguish
the central object or star, the pressure is calculated via a
barotropic equation of state (EOS) 
\begin{align}
  P = \ciso^2\rho^{\gamma_1}\left[1 +
    \left(\frac{\rho}{\rho_*}\right)^{\gamma_2 - \gamma_1}\right], 
\end{align} 
where $\gamma_1=1.00001,\,\gamma_2=5/3$ and $\rho_*$ is a fixed
transition density defined below.

This EOS  mimics star formation by halting gravitational collapse.  
For $\rho\ll\rho_*$, the fluid collapses isothermally. As the collapse
proceeds, the cloud's central density increases. When $\rho\gg\rho_*$,
$P\propto\rho^{5/3}$ and further collapse is prevented by the
increased pressure. The central object is then effectively a polytrope
with polytropic index $n=3/2$. Because it has finite volume, it can
deform. The resulting non-axisymmetric object may then be spun down
due to gravitational torques from the external, lower density disc.

\subsection{Initial conditions}
The system is initially an isothermal sphere of radius $r_c$. For 
$r\geq r_*\equiv qr_c$, where $r$ is the spherical radius and $q$ is a  
dimensionless parameter, the initial density is 
\begin{align}\label{initial_density} 
  \rho_0(r) = \frac{A\ciso^2}{4\pi G r^2},\quad r\geq r_*. 
\end{align}
The dimensionless parameter $A$ relates to the accretion rate $\xi$, 
and we use tabulated values of $A$-$\xi$ pairs from \cite{shu77}.  
The spherical region $r\leq r_*$ is designated as the initial star.   
We set $\rho_*=\rho_0(r_*)$ and $\rho_0(r<r_*)=\rho_*$.  

The cloud is initialised with an azimuthal velocity  
\begin{align}\label{initial_rotation} 
  v_\phi = 2A\ciso h\times
  \begin{cases} 
    R/r_* & R \leq r_* \\
    1 & R > r_*, 
  \end{cases} 
\end{align} 
where $R$ is the cylindrical radius. The initial star 
has solid body rotation and is below break-up speed at its equatorial
plane but faster than the rest of the core. The latter may bias  
angular momentum loss from the star to the disc, because if the star 
becomes triaxial, the pattern speed of the stellar
potential felt by the disc may be naturally higher than the disc
rotation frequency, exciting trailing spiral density waves in the
disc.

% Apart from the small region $r\leq qr_c$, the initial condition here is
% identical to that of \cite{kratter10}. The collapse evolution of cores with
% density and rotation profiles stated above have already been
% thoroughly investigated by \citeauthor{kratter10}. They found that
% the constant $v_\phi$  core leads to roughly constant $\Gamma,\,\xi$
% during the simulation. This property motivated us to follow
% \citeauthor{kratter10}'s setup. However, we stress that for our
% problem, $\Gamma,\xi$ serve as 
% initialisation parameters. We shall see that our systems consists of
% an non-axisymmetric central mass with a non-axisymmetric disc, with
% possible angular momentum exchange between the two. This situation is
% clearly very different to the axisymmetric problem these parameters
% describe.
  
  Apart from the small region $r \leq q r_c$, the initial conditions here
  are identical to that of \cite{kratter10}. Our motivation for this
  setup is the same as in that paper: this setup produces a collapse in
  which the dimensionless parameters $\xi$ and $\Gamma$ remain fixed,
  and thus produce a particularly clean numerical experiment for
  studying behaviour as a function of $\xi$ and $\Gamma$. A real
  protostellar core, of course, will not follow this simple structure,
  and $\xi$ and $\Gamma$ will vary as the core accretes. However,
  this
  variation will generally be on time scales that are long compared to
  the disc orbital period \cite[e.g. see][]{kratter08,offner10}
  and we may therefore, to first order, think of the system as
  simply proceeding through a series of quasi-equilibrium states with
  fixed $\xi$ and $\Gamma$. Thus experiments at given $\xi$ and $\Gamma$
  provide valuable insight. 
  
  It should be noted, though, that in the idealised collapse described by
  constant $\xi$, $\Gamma$, the central star is a point mass that can neither gain nor
  lose angular momentum, whereas in our problem it can freely exchange angular
  momentum with the disc. This effect is neglected in simple collapse models, so our 
  cores may evolve away from it. With this in mind, it is best to regard $\xi,\Gamma$ as  
  initialisation parameters.

\subsection{Dimensionality and physical scales}\label{dimensionless}   
As in the self-similar collapse of \cite{shu77} and in 
  \cite{kratter10},  the problem we have specified is fully
dimensionless, and so we report our results in terms of
non-dimensional numbers. In the evolutionary plots shown below,
  time is scaled by $\tau \equiv 2\pi q(r_c/\ciso)\sqrt{3/A}$, which
  is the Keplerian orbital period at $r=qr_c$.

  Because our problem is dimensionless, we can scale our simulations
  to apply to a range of physical systems, in the context of low and high
  mass star formation (and possibly even gaseous planet formation).
  However, there is one caveat: the dynamic range we are able to achieve
  in our simulation is much less than the dynamic range involved in the
  formation of a real star ($r\sim 10^{11}\mathrm{cm}$) from a real protostellar
  core ($r\sim 10^{17}\mathrm{cm}$). Even with adaptive mesh
  refinement, we achieve a dynamic range of $\sim 10^4$, not $\sim
  10^6$. If one wishes to scale our  dimensionless experiments to
  physical scales, one may think of doing so in two ways
  that give dynamic range comparable to what we achieve.   
  
  First, one could envision that we are studying the collapse
  of a protostellar core leading up to the formation of the first
  hydrostatic core that is $\sim 5\mathrm{AU}$ in size \citep{masunaga98}.
  Alternately, one could envision that our central object is the size of
  a star, but that we are simulating the evolution only of the inner $
  \sim 10^{15}$ cm of the core that surrounds it. However, there is no
  evidence that the physical effects we identify depend in the slightest
  on the dynamic range of the simulations.

\subsection{Numerics}\label{numerics}
The hydrodynamic equations are evolved using the Godunov-type ORION  
code \citep{truelove98,klein99,fisher02} in Cartesian co-ordinates
$(x,y,z)$ . The computational domain is a cube of length
$L=4r_c$. ORION offers adaptive mesh refinement, a key advantage for
the multi-scale flow considered here. We do not make use of ORION's
sink particle or radiative transfer capabilities in this study. We use
a base grid of $128^3$ with 6 levels of refinement, giving the highest
effective resolution of $8192^3$. 

A characteristic length-scale for self-gravitating problems is the
Jeans length, 
\begin{align}\label{jeanslength}
  \lambda_J = c_s\sqrt{\frac{\pi}{G\rho}},
\end{align}
where $c_s= \sqrt{dP/d\rho}$ is the density-dependent sound-speed. For
a grid spacing $\delta x$, a measure of resolution is $\delta
x/\lambda_J$.  Our EOS gives 
\begin{align}
  \frac{\lambda_J}{\delta x} = \frac{\ciso}{\delta
    x}\sqrt{\frac{\pi}{G\rho}} 
  \times\left[1+\frac{5}{3}\left(\frac{\rho}{\rho_*}\right)^{2/3}\right]^{1/2}.  
\end{align}
For example, if $\rho/\rho_* \sim 10$, then the linear resolution is a
factor of $\sim 3$ better than it would be for gas of the same density
obeying an isothermal EOS. 

We use a Jeans number $N_J=8$ to define the maximum resolvable density
$\rho_J$ 
\begin{align}\label{jeansnumber}
  \rho_J \equiv \left(\frac{\ciso}{N_J\delta x}\sqrt{\frac{\pi}{G}}\right)^2,
\end{align}
and we refine if $\rho>\rho_J$ at each grid level. Note the above is 
the isothermal Jeans density, which is smaller than the true Jeans
density based on our EOS. Thus, using $\rho_J$ as defined above is a
conservative approach.  To ensure the star-disc interface is resolved,
we also refine if $\rho > 0.5\rho_*$.  We also refine to the highest
level within the theoretical disc radius $R_k$, and within one scale-height in
$z$.

\section{Diagnostics}\label{diagnostics}
In this section we define the diagnostic tools used to interpret our 
results. Measurements of the star are summarised in Table
\ref{star_defs}, where integrals are taken over the region
$\rho\geq\rho_*$ ($dV$ is the volume element); symbols preceded by
$\Delta$ are relative to the star; `KE' and `PE' stand for kinetic and
potential energies, respectively. Note that $\Phi$ is the
gravitational potential due to all of the fluid.

We regard fluid with $\rho \geq \rho_*$ as stellar. The star
has position and velocity $(\bm{x}_*,\bm{v}_*)$. The stellar spin
angular momentum is defined with position and velocities of fluid
elements with respect to $(\bm{x}_*,\bm{v}_*)$. The stellar rotation
radius $S_*$ is defined from its moment of inertia and is used to
define stellar spin and break-up frequencies. $S_*$ is not to be
confused with the stellar surface where $\rho = \rho_*$, which is
typically larger than $S_*$ and generally non-axisymmetric. We use
\ssize to denote the average radius in the star's equatorial plane
where $\rho = \rho_*$.

The evolution of star's angular momenta is strongly influenced by the
fluid external to the star. The properties of the fluid are measured
in cylindrical co-ordinates $(R,\phi,z)$ centred on the star with
velocities relative to $\bm{v}_*$. We take the vertical direction of
the cylindrical co-ordinates to be parallel to the vertical ($z$)
direction of the inertial frame, assumed to be aligned with the
stellar spin axis. The disc mass $M_d$ is defined to be the difference
between the total mass within a cylinder of radius $R_k$, thickness
$2hR_k$, centred about the star, and $M_*$. The disc mass is only
defined when the mass within this cylinder exceeds $M_*$.

For simplicity, we integrate the external fluid vertically over a slab 
of constant thickness and work with surface density $\Sigma\equiv \int
\rho dz$ and vertically-averaged velocity $\bm{U}$, where
\begin{align}
  \bm{U} = \frac{1}{\Sigma}\int\rho(\bm{v} - \bm{v_*})dz.
\end{align}
Our results are insensitive to the extent of vertical integration, 
and in fact also insensitive to whether we use relative velocities or
those in the inertial frame. 

Non-axisymmetric modes in the surface density of the external fluid 
can be found via Fourier analysis defined by
\begin{align}
  a_m(R) \equiv \int \Sigma(R,\phi) \exp{(-im\phi)} d\phi,
\end{align}
where $a_m(R)$ is the radius-dependent amplitude and $m$ is the
azimuthal wave-number. The integrated amplitude $C_m = \int a_m dR$ is
used as a global measure of mode amplitudes. This integration excludes
the star.

The relevant angular momentum fluxes for this problem are:
\begin{align}
  &F_A = R\Sigma U_\phi U_R \\
  &F_R = R\Sigma \delta U_\phi \delta U_R,\\
  &F_G = \int dz \partial_R \Phi \partial_\phi \Phi/4\pi G. 
\end{align}
$F_A$ is the flux associated with large-scale advection and $F_R$ is
the flux due to Reynolds stresses, where $\delta$ here denotes
deviations from azimuthally-averaged values. $F_G$ is the angular
momentum flux associated with self-gravitational torques
\citep{lynden-bell72}. These fluxes can be converted into $\alpha$
viscosities  denoted $\alpha_A,\, \alpha_R,\,\alpha_G$ e.g.,
\begin{align}
  \alpha_G =F_G \left/ \left\langle
      R\left|\frac{d\ln{\Omega}}{d\ln{R}}\right|\Sigma\ciso^2\right\rangle_\phi\right.,  
\end{align}
and similarly for $\alpha_A,\,\alpha_R$. $\Omega\equiv U_\phi/R$ is
the angular velocity and $\langle\cdot\rangle_\phi$ denotes an azimuthal
average.  

It is important to regard the $\alpha$'s above simply as
non-dimensionalised fluxes rather than the viscosity coefficient in
accretion disc models \citep[][which would impose $ 0 \leq \alpha \la
1$]{ss73}. A positive flux out of the star indicates spin angular
momentum loss. To ensure that this angular momentum loss is physical
rather than numerical (since ORION does not exactly conserve angular
momentum), we can compare the physical $\alpha$'s we measure to the
effective numerical viscosity, denoted by $\alpha_N$, measured 
for the ORION code by \cite{krumholz04}.

It will also be useful to consider mass accretion
  rates. To measure the $\xi$ parameter from our simulations, we define
  \begin{align}
    \xi_\mathrm{sim} \equiv \frac{G}{\ciso^3}\frac{d}{dt}\left(M_* +
      M_d\right). 
  \end{align}
  Although this is the most natural definition, because frequent
  simulation output is not practical due to the large data sizes
  involved, the numerical time derivative only gives an
  estimate. Nevertheless, we find that the mass accretion rate is consistent
  with the behaviour of stellar spin. 
% We define $M_\mathrm{dot}$ as
%  \begin{align}
%    M_\mathrm{dot} \equiv -2\pi R \langle \Sigma U_R\rangle_\phi,
%  \end{align}
%  and consider the non-dimensional flux
%  $GM_\mathrm{dot}/\ciso^3$. This is the mass flux in the cylindrical
%  radial direction, and should be distinguished from the initialisation 
%  parameter $\xi$, which describes the rate of increase of
%  mass. 
% 

\begin{table}
 \caption{Definition of stellar measurements.}
 \label{star_defs}
 \begin{tabular}{@{}lll}
  \hline
  Name & Symbol & Definition \\
  \hline
  Mass & $M_*$ & $\int \rho dV$ \\
  Position & $\bm{x}_*$ & $ \int \bm{x}\rho dV/M_*$ \\
  Velocity & $\bm{v}_*$ & $ \int \bm{v}\rho dV/M_*$ \\
  Rotation radius & $S_*$ & $\left\{\int \left[\Delta x ^2 + \Delta y ^2\right]\rho dV/M_*\right\}^{1/2} $ \\
  Orbital ang. mom. & $j_\mathrm{o}$ & $\bm{x}_* \wedge \bm{v}_*\cdot\hat{\bm{z}}$ \\
  Spin ang. mom. & $j_\mathrm{s}$ &
  $\int \rho \Delta \bm{x} \wedge \Delta\bm{v} \cdot\hat{\bm{z}} dV /M_*$ \\
  Spin freq. & $\Omega_\mathrm{s}$ & $j_\mathrm{s}/S_*^2$ \\
  Break-up freq. & $\Omega_\mathrm{b}$ & $\sqrt{GM_*/S_*^3}$ \\
  KE-to-PE &$T/|W|$ &$\int \rho|\Delta \bm{v}|^2 dV/|\int \rho \Phi dV|$ \\
  \hline
\end{tabular}
\end{table}

\section{Results}\label{results}
We study the evolution of the star's spin in two cases. In Case 1, we
consider initial conditions leading to a large disc-to-star mass ratio.  
For our comparison (Case 2) run we consider initial conditions
which lead to a smaller disc-to-star mass ratio. We shall see
that on the overall simulated timescale, the massive disc (Case 1) is
less efficient at spinning down the star than its lower mass
counterpart (Case 2) due to different azimuthal spiral modes
dominating the angular momentum transport.

  Before discussing our two cases, we should mention an important
constraint on our parameter choices. \cite{kratter10} have established
the ranges in parameter space for which disc fragmentation is or 
is not expected. We limit ourselves to values of $
\xi$ and $\Gamma$ such that the disc will be non-fragmenting within
the simulation. We do so in
order to render the numerical experiment as clean and easy to
interpret as possible. 

Massive stars form with high accretion rates that do tend to
produce gravitationally-unstable, fragmenting discs
\citep{krumholz07,krumholz09,kratter06,kratter08,peters10,hennebelle11}. 
However,  protostellar discs are heated by the central star and by viscous
dissipation, both of which increase at smaller radii, so that fragmentation generally occurs only outside $\sim
100\mathrm{AU}$. Since it is the angular momentum flux out of the star to
the disc that counts, it is the inner disc that is important for
star-disc angular momentum exchange. This means that our choice to
limit ourselves to the non-fragmenting part of parameter space does
not prevent us from applying our models to massive protostars and
their discs.     

%  Before discussing our two cases, we should mention an important
%  constraint on our parameter choices. We limit ourselves to values of $
%  \xi$ and $\Gamma$ such that the disc will be non-fragmenting within
%  the simulation (\cite{kratter10} established the ranges in parameter
%  space for which fragmentation is or is not expected.) We do so in
%  order to render the numerical experiment as clean and easy to
%  interpret as possible. However, this raises the question of the extent
%  to which our models can apply to the problem of massive star
%  formation. Massive stars form with high accretion rates that tend to
%  produce gravitationally-unstable, fragmenting discs
%  \citep{krumholz07,krumholz09,kratter06,kratter08,peters10,hennebelle11}. 
%  
%  However, this is less of a problem than it appears. Because massive
%  protostellar discs are heated by the central star and by viscous
%  dissipation, both of which increase at smaller radii, the inner
%  parts of massive protostellar discs are not gravitationally unstable
%  or fragmentation; fragmentation generally occurs only outside $\sim
%  100$ AU. Since it is the angular momentum flux out of the star to
%  the disc that counts, it is the inner disc that is important for
%  star-disc angular momentum exchange. This means that our choice to
%  limit ourselves to the non-fragmenting part of parameter space does
%  not prevent us from applying our models to massive protostars and
%  their discs.     

\subsection{Case 1}\label{case1}
This case has parameters $\xi = 5.58$ ($A=4.0$), $ h=0.1$ and $
q=0.005$. We start measurements after the star is sufficiently
resolved at the finest grid level. The radius \ssize is typically
resolved by at least $20$ cells.

\subsubsection{Self-limited spin up}\label{self_limited1} 
\begin{figure*}
  \centering
  \includegraphics[scale=.67 ,clip=true,trim=0cm 1.3cm 0.48cm
  0cm]{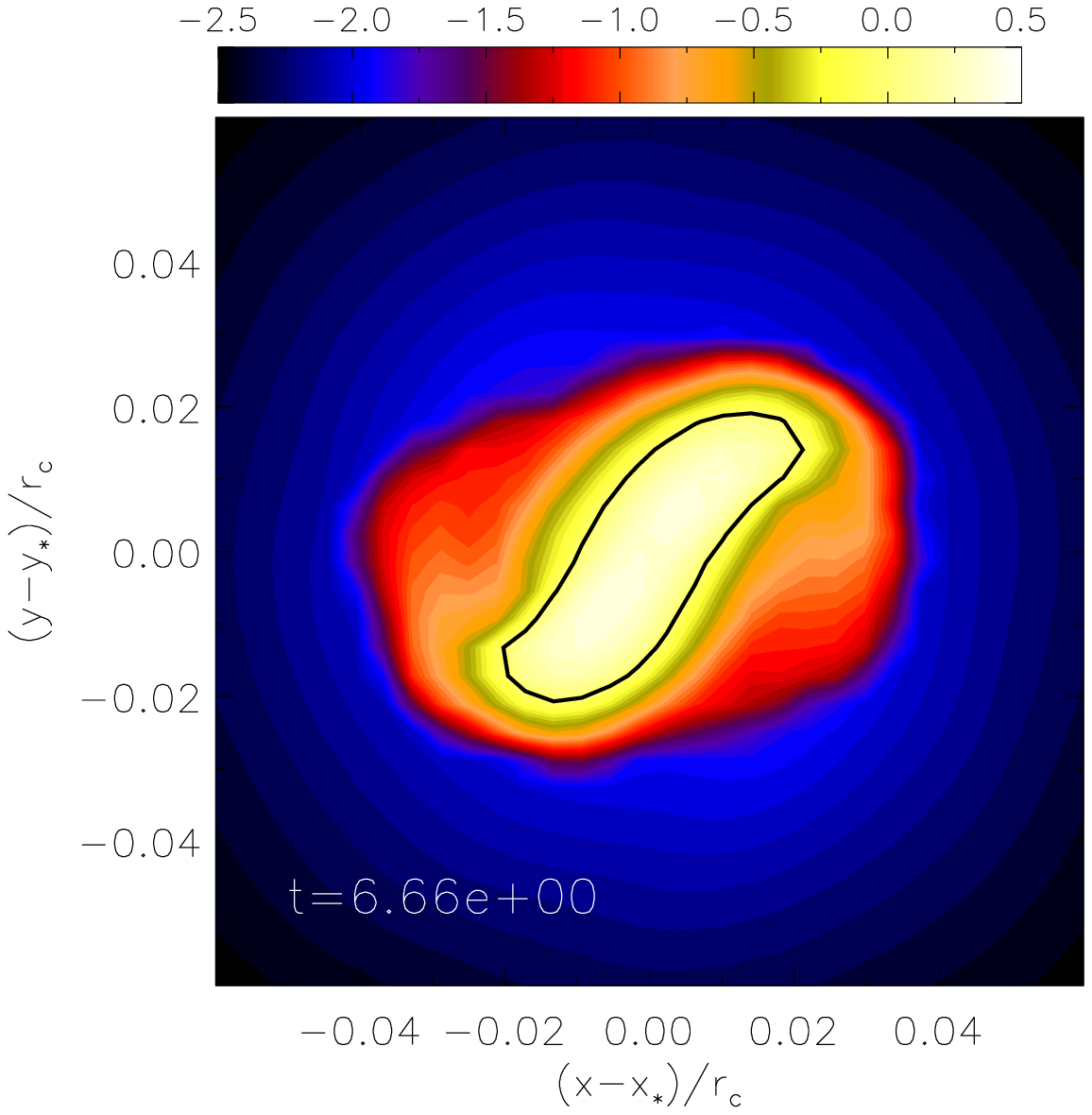}\includegraphics[scale=.67,clip=true,trim=2.1cm 1.3cm 0.45cm 0.0cm]{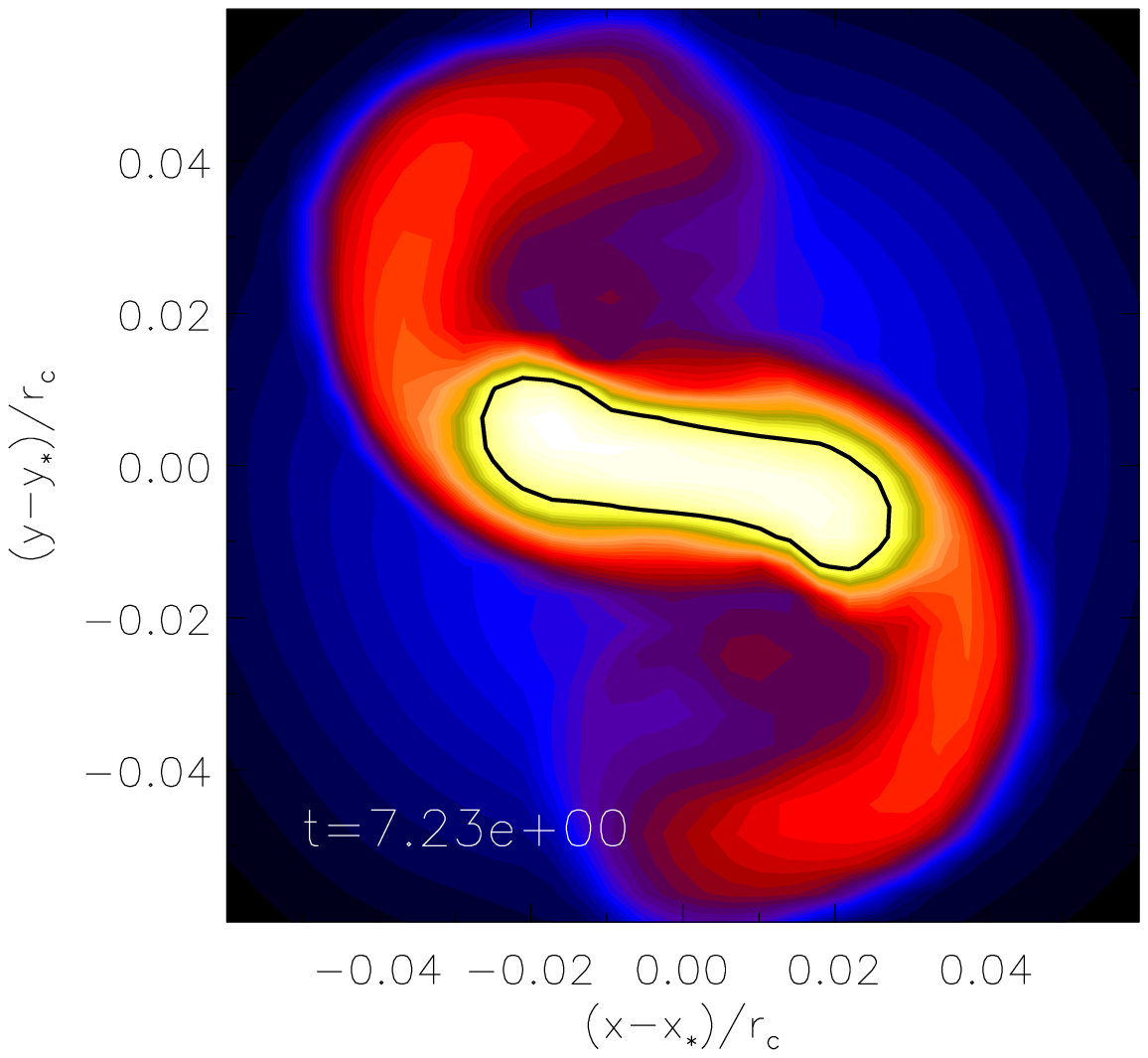}       
  \includegraphics[scale=.67 ,clip=true,trim=0cm 0cm 0.48cm
  1.35cm]{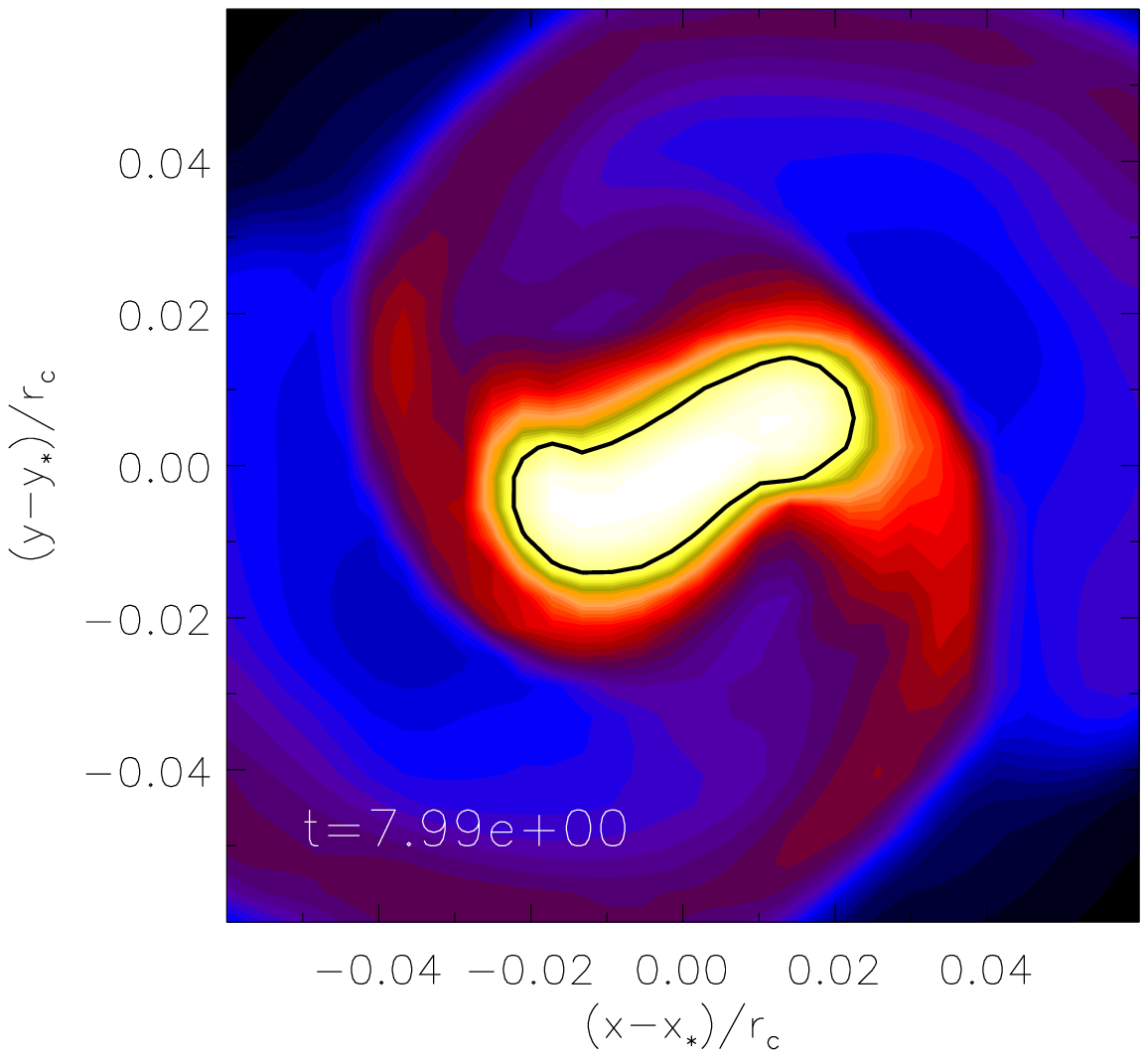}\includegraphics[scale=.67,clip=true,trim=2.1cm 0cm 0.45cm 1.35cm]{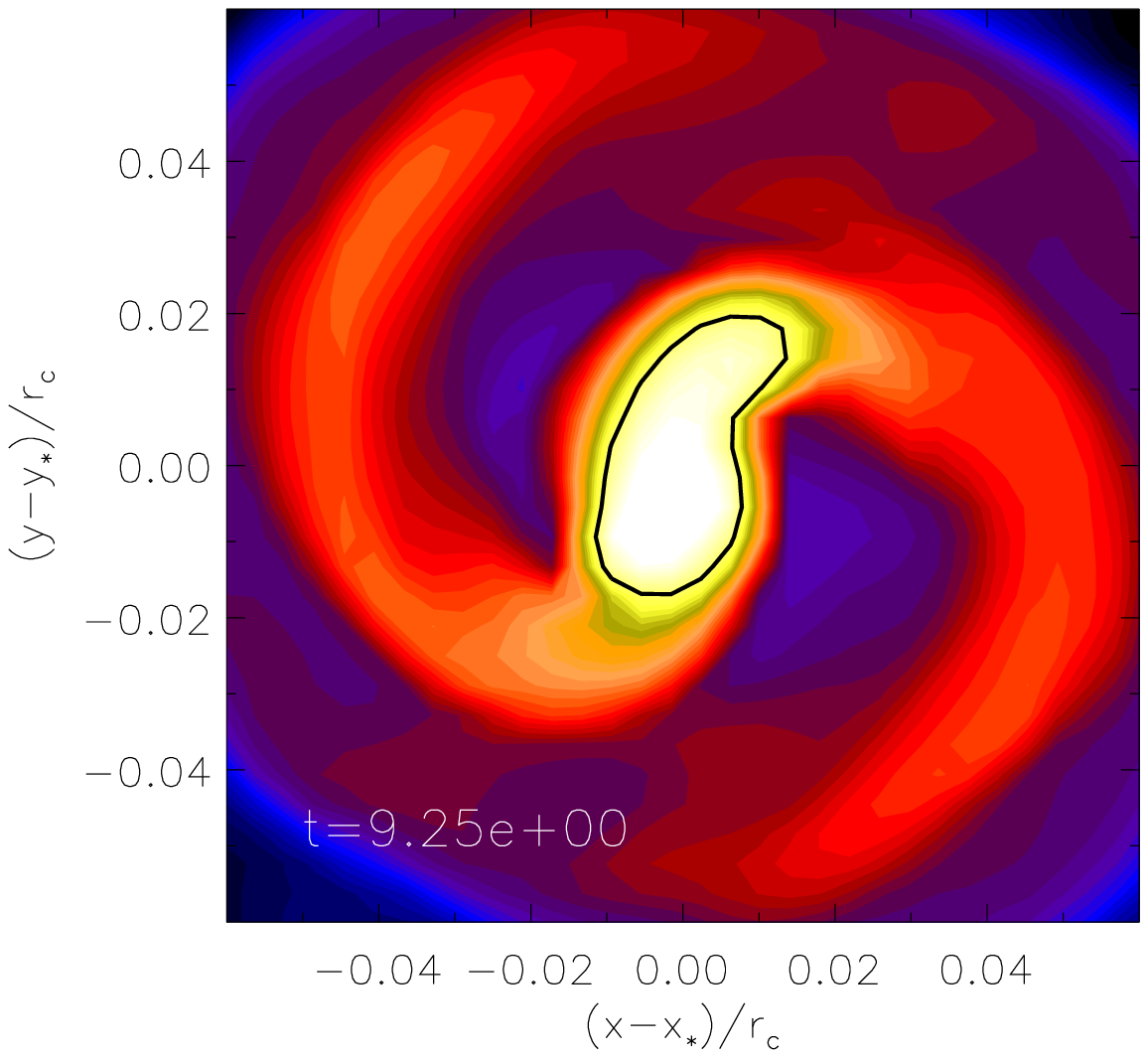} 
  \caption{Case 1: contours of $\log{(\rho/\rho_*)}$ in the star's
    equatorial plane at four snapshots during the initial collapse.   
    Thick lines delineate $\rho=\rho_*$.  
    \label{case1HR_overview_snapshots}} 
\end{figure*}

\begin{figure}
  \centering
  \includegraphics[width=0.95\linewidth]{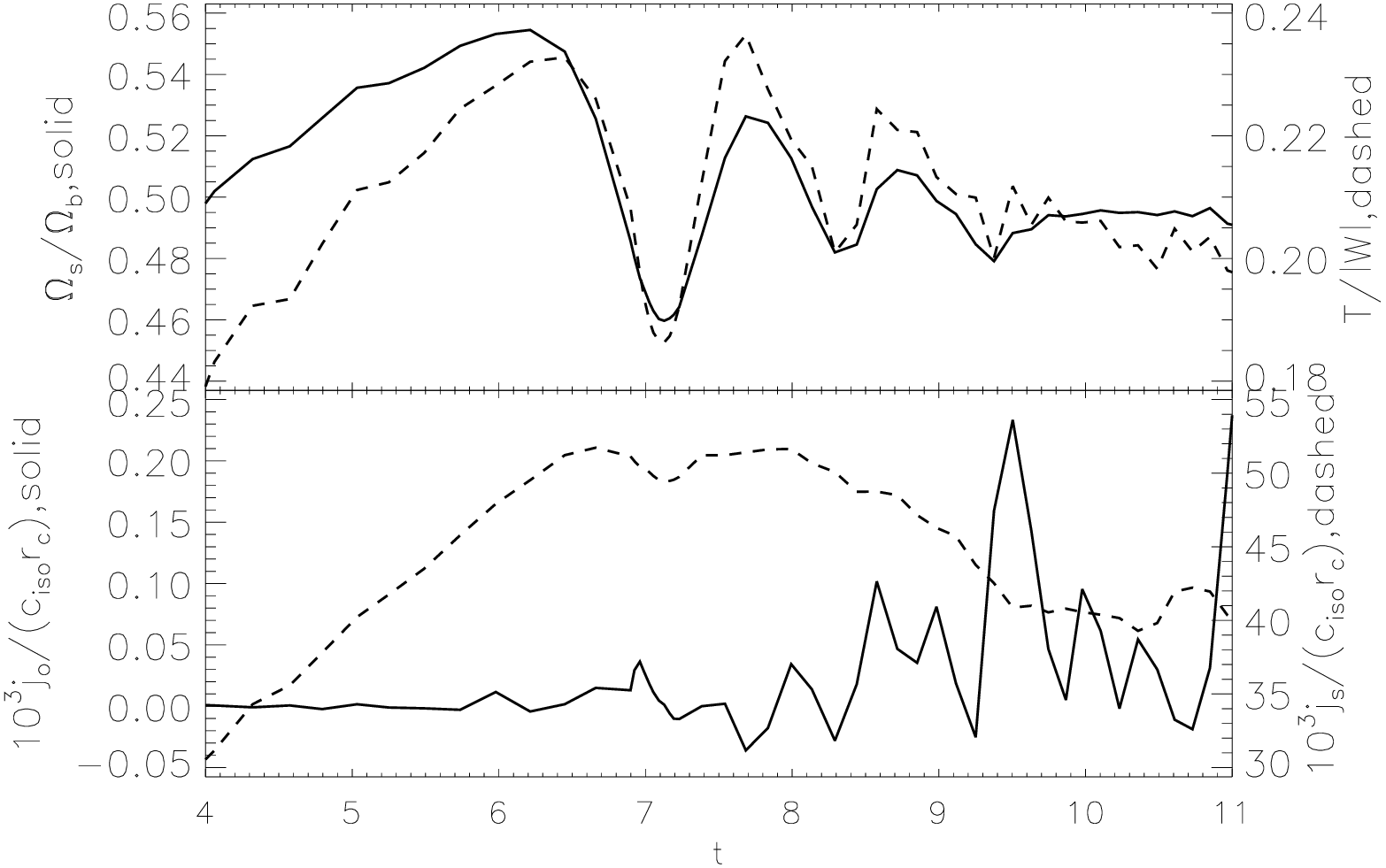} 
  \caption{Evolution of stellar properties for Case 1 during the
    initial collapse phase. The top panel shows the spin to break-up
    frequency ratio (solid) and the  ratio of kinetic to potential
    energy (dashed). The bottom panel shows the orbital angular
    momentum (solid) and spin angular momentum (dashed). Note the 
    different scales used on the left and right vertical axes.
    \label{case1HR_overview_evolution}}
\end{figure}

We first describe the early evolution $4 \leq t\leq 11$, when the disc 
mass is insignificant compared to the stellar
mass. Fig. \ref{case1HR_overview_snapshots} and
Fig. \ref{case1HR_overview_evolution} show snapshots of the density
field and evolution of stellar properties, respectively.

At $t\la 6.2$, the star spins up with a rapid increase in  $\jspin,\,
\spin/\brk$ and $T/|W|$ as it accretes material and angular
momentum. Because $\Omega_s=j_s/S_*^2$,  spin-up can occur because of
decreasing $S_*$, but our EOS inhibits further collapse when
$\rho\gg\rho_*$. Hence, the spin-up is due to accretion, as expected
for initial collapse. As it spins up, the star deforms into a bar-like
object (Fig. \ref{case1HR_overview_snapshots}, $t=6.66$). The
bar-shaped star exerts a positive gravitational torque on the
surrounding material because it spins faster than the surrounding
material, producing spiral arms in the latter
(Fig. \ref{case1HR_overview_snapshots}, $t=7.23$). This counteracts
the increase in stellar spin angular momentum due to the accretion of
material, resulting in an approximately constant $\jspin$ during $6.2
\la t\la 7.7$.

From $t=8$ to  $t=10$ spiral arms are always present in the
external fluid (e.g. Fig. \ref{case1HR_overview_snapshots},
$t=9.25$) and Fig. \ref{case1HR_overview_evolution} shows that
$\jspin$ is decreasing during this time interval. This indicates spin
angular momentum loss due to negative torque exerted on the star by
the prominent spiral arms. Thus the initial spin-up is limited by the
increasing spin-down torque from the external fluid as the
star becomes deformed into a non-axisymmetric, $m=2$ object. 
Notice in Fig. \ref{case1HR_overview_evolution} that
$|\jorb|\ll|\jspin|$ so the star is essentially fixed in the $x-y$
plane in the inertial frame. However, note that the increase in 
$\jorb$ at $t=9.4$  coincides with when $\jspin$ and $\spin/\brk$ stop
decreasing. This suggests orbital motion hinders spin-down.

\begin{figure}
  \centering
  \includegraphics[width=\linewidth]{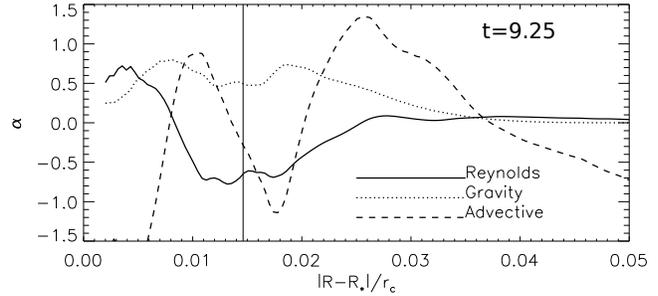}
  \caption{Case 1 early phase spin-down: non-dimensionalised,
    azimuthally averaged angular momentum fluxes due to the
    large-scale advection, self-gravity and Reynolds stresses.
    $|R-R_*|$ denotes the cylindrical distance away from the star. The
    vertical line is \ssize$=0.0146r_c$. This snapshot corresponds to
 the    bottom 
    right contour plot in Fig. \ref{case1HR_overview_snapshots}. 
    \label{case1HR_stardisc}}
\end{figure}

Fig. \ref{case1HR_stardisc} compares the angular momentum fluxes at $t=9.25$,
when $\jspin$ is decreasing\footnote{Defining the external fluid
  relative to the star implies an 
  indirect potential associated with stellar motion, but this vanishes
  when performing azimuthal averages.}. 
The rise of gravity flux towards the average stellar surface, to
$\alpha_G\sim 0.5$ at \ssize, implies some angular momentum loss due to
gravitational torques. The figure shows that $\alpha_A<0$ around
\ssize which indicates that large-scale advection is spinning up the star,
but $\alpha_G+\alpha_A\sim 0.2>0$, implying that gravitational
spin-down torques outweigh the effect of accretion, thereby preventing
spin-up and leading to spin-down. The numerical viscosity is
estimated to be $\alpha_N\sim 0.03$ near $\rho\simeq\rho_*$ and is
therefore negligible.  

%%%%%%%% DO NOT DELETE%%%%%%%%%%%%%%%%%
%radius is     0.014615042
%alpha_g       1.5566161
%alpha_t     -0.17920907 (turbulent part)
%alpha_t      0.63884286  (large scale advection)

\begin{figure} 
  \centering 
  \includegraphics[width=\linewidth]{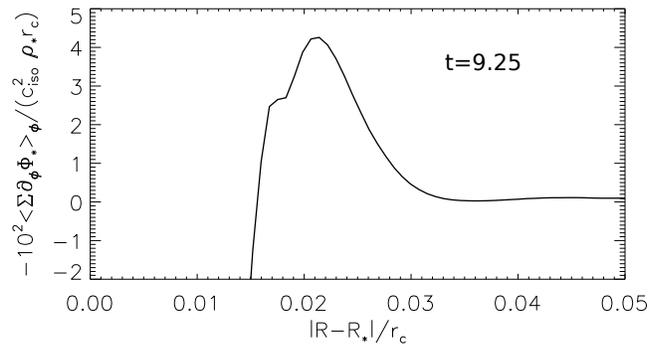} 
  \caption{Case 1 early phase spin-down. Azimuthally averaged, 
    non-dimensionalised torque exerted by the star on its
    surroundings. The region $\la 0.015r_c$ 
    corresponds to the star.  
    \label{case1HR_stardisc_torque}}
\end{figure}

We also explicitly solve for the gravitational potential $\Phi_*$  
associated with $\rho>\rho_*$ and find a positive torque exerted by 
the star on the external fluid,  as shown in
Fig. \ref{case1HR_stardisc_torque}. This torque is concentrated near
\ssize. Together with the contour plots in
Fig. \ref{case1HR_overview_evolution}, this is consistent with
prominent $m=2$ spirals in the density field being responsible for
spinning down the star.    

\subsubsection{Long term evolution}
\begin{figure}
  \centering
  \includegraphics[width=\linewidth]{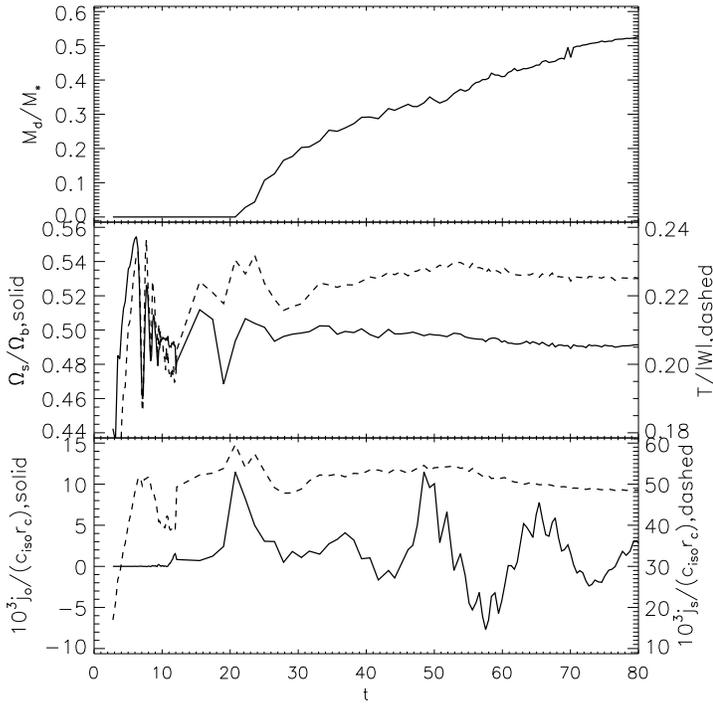}
  \caption{Case 1: evolution of the disc-to-star mass ratio (top) and the same 
    stellar properties shown in Fig. \ref{case1HR_overview_evolution}
    (middle, bottom) over the entire simulation.  Note the different
    scales used for the left and right vertical axes.   
    \label{case1HR_star}}
\end{figure}

We ran Case 1 until the stellar spin approached a steady state. 
Its evolution is summarised in Fig. \ref{case1HR_star}. 
Notice that the early phase discussed previously, when most variation
in stellar spin occurs, coincides with the time when the predicted
disc is undefined.

Between $ 11 \la t \la 26$,  $\spin/\brk$ is highly variable but
settles to $\simeq0.5$ at $t=34$. Between $39 \la t \la
74$ it decreases at a negligible rate compared to $7.7\la t\la
11$. Most of the variation in $T/|W|$ also diminishes after $t=26$.   
Notice that $\spin/\brk$ and $T/|W|$ become relatively constant 
as $M_d/M_*$ increases.

The stellar spin angular momentum behaves similarly. There is no
appreciable change in $\jspin$ relative to $t\la 11$ after disc
formation. However, between $11 \la t\la 20$, the orbital angular
momentum increases by two orders of magnitude (comparing $\jorb$ in
Fig. \ref{case1HR_overview_evolution} and Fig. \ref{case1HR_star})
from essentially zero 
to values comparable to $\jspin$: at 
$t\simeq 20$, $\jorb\simeq 0.056$ and $\jspin\simeq
0.011$.

As $M_d / M_*$ increases
from $\sim 0.3$ to $\sim 0.5$, the orbital angular momentum $j_o$
undergoes large oscillations, which are comparable in magnitude to $j_s$.
Indeed, we found that the star exhibits complex orbital motion, and its
displacement from the box centre can be $\sim 0.022r_c$, larger than
its size (\ssize $\simeq 0.0146r_c$). Significant orbital
motion coincides with disc formation.

\begin{figure}
  \centering
 \includegraphics[width=\linewidth]{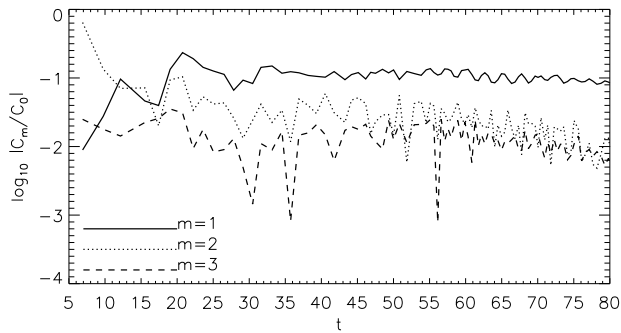}
  \caption{Case 1: evolution of non-axisymmetric
    modes in the surface density of a region of cylindrical 
    radius $0.25r_c$ and thickness $0.025r_c$ about the
    star. Fourier amplitudes were integrated from a cylindrical radius 
    of $0.02r_c$ to $0.25r_c$ away from the star, then normalised 
    by the axisymmetric amplitude.   
    \label{case1HR_fft_time}}
\end{figure}

The evolutionary plots suggest that the disc %in fact
inhibits changes to stellar spin. Instead,
its interaction with the disc leads to orbital motion. This is
consistent with a Fourier analysis of the surface density shown in 
Fig. \ref{case1HR_fft_time} \footnote{As the predicted disc does not develop
until $t\simeq20$, strictly speaking amplitudes prior to this
time are not disc modes but simply that of the external medium.}. 
During the self-limited spin-up phase ($t<11$), the $m=2$ mode is
dominant in the external medium to the star. The $m=1$ mode over-takes
$m=2$ at $t\sim 15$, which is when $\jorb$ begins to increase. After
$t\simeq 20$, the disc forms a strong $m=1$ asymmetry.  
The dominance of the $m=1$ mode coincides with limited evolution
in stellar spin, but large amplitudes in orbital motion.

Early theoretical work shows that an $m=1$ lopsided over-density
causes stellar orbital motion about the system's centre of mass 
\citep{adams89, heemskerk92}. In those studies
the star is treated as a point mass, so star-disc torques can
\emph{only} affect the star's orbital motion. Although we model the 
star as a finite-sized polytrope, and thus have an 
associated spin angular momentum, the angular momentum exchange between
the star and the $m=1$-dominated disc is still in orbital angular
momentum. Changes to the spin angular momentum are apparently inhibited. 

\begin{figure} 
    \centering 
    \includegraphics[width=\linewidth]{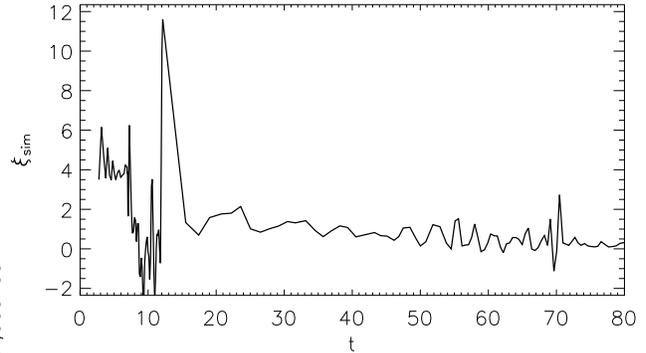}
    \caption{Case 1: non-dimensionalised mass accretion rate. The
      large peak at $t\simeq 12$ may be an artifact from numerical
      time derivatives. 
      \label{case1HR_mass_acc}}
  \end{figure}

  Fig. \ref{case1HR_mass_acc} shows the evolution of the dimensionless
  mass accretion rate. During the very early stages ($t\la 7$),
  i.e. the initial spin-up phase, $\xi_\mathrm{sim}\sim 4$ and roughly
  constant (recall the initialisation parameter
  $\xi=5.58$). $\xi_\mathrm{sim} < \xi $ since the 
  core has angular momentum so spherically symmetric collapse cannot
  occur. The sharp drop in $\xi_\mathrm{sim}$ in $7 \la t\la 10$
  coincides with  spin-down. As the star loses angular momentum to the
  immediate surroundings, the latter gains angular momentum, inhibiting accretion. 
  However, once the
  star-on-disc torque is reduced (because the star has spun down and
  become more axisymmetric), % less non-axisymmetric), 
  material can again fall in and
  $\xi_\mathrm{sim}$ rises ($10 \la t \la 20$). When the stellar spin
  is in steady state $\xi_\mathrm{sim} < \xi$, but remains
  of order unity. The lower $\xi_\mathrm{sim} $ may result from
  the fact that at later times, the disc is lopsided and is not well
  described by accretion from a spherical cloud onto an axisymmetric
  disc.  
  
%   Fig. \ref{case1HR_mass_flux} shows the mass flux at three
%   snapshorts. At $t= 6.66$, the mass fluxes are smaller
%   than $\xi$ but similar magnitude (although the mass
%   flux and $\xi$ have different meanings). Again, this is probably 
%   because of reduced accretion in the direction perpendicular to the 
%   object's equitorial plane. From
%   Fig. \ref{case1HR_overview_snapshots} we see that the central object
%   extends to $0.02r_c$ in size, and Fig. \ref{case1HR_mass_flux}
%   shows that for $t\leq 10$ the region just outside $0.02r_c$ has a
%   positive gradient, indicating that the local mass increases with
%   time. The mass flux is much reduced at
%   longer time-scales ($t=68.7$), with more complicated features
%   towards the central object. The reduced flux correlates with the
%   formation of a lopsided disc orbiting the central object, in which
%   case one expects small radial flow.   

%   \begin{figure} 
%     \centering 
%     \includegraphics[width=\linewidth]{figs/case3_mdot_compare.ps}
%     \caption{Case 1: non-dimensionalised mass flux. 
%       \label{case1HR_mass_flux}}
%   \end{figure}  

\subsubsection{Structure and angular momentum transport at late stages}
\begin{figure}
  \centering
  \includegraphics[width=\linewidth]{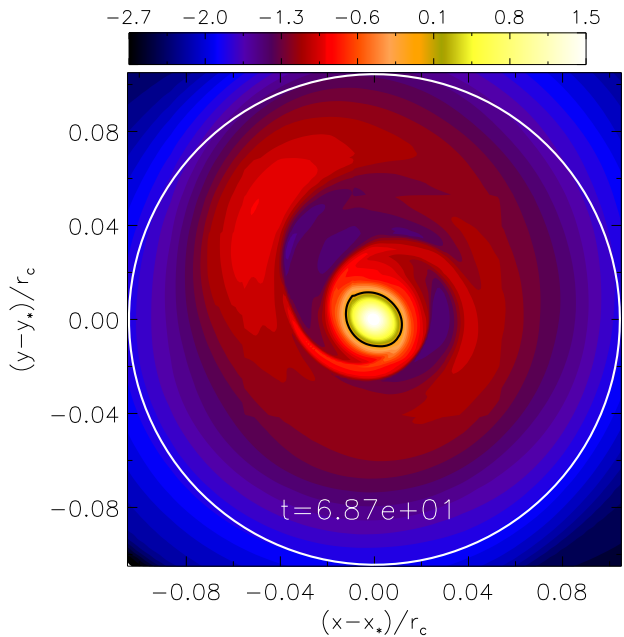}
  \caption{Case 1 towards the end of the simulation. The 
      colour-bar indicates $\log{\rho/\rho_*}$ in the star's
    equatorial plane. The inner black line indicates $\rho=\rho_*$
      and the outer white circle indicates the disc edge $R_k$.
    \label{case1HR_disc}} 
\end{figure}

\begin{figure}
  \centering
 \includegraphics[width=\linewidth]{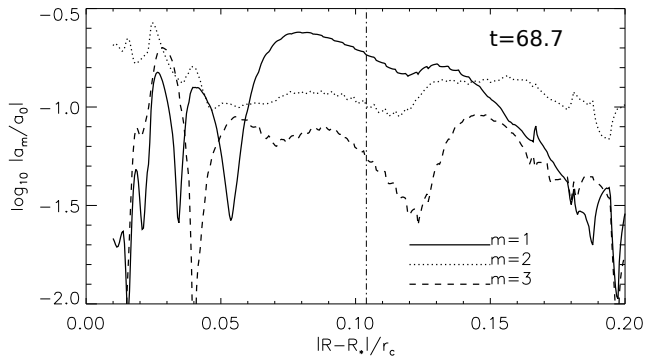}    
  \caption{Case 1 towards the end of the simulation: Fourier 
    amplitudes of non-axisymmetric modes in surface density, 
    corresponding to the snapshot in Fig. \ref{case1HR_disc}.   
    The vertical line indicates the predicted disc radius $R_k$.  
    \label{case1HR_disc_modes}}
\end{figure}

In order to understand why spin evolution is inhibited on timescales 
beyond the early phase, we analyse the disc structure
while $\jspin$ and $\spin/\brk$ remain approximately
constant. Fig. \ref{case1HR_disc} and Fig. \ref{case1HR_disc_modes}
respectively show the density field in real and Fourier space at $t=68.7$. 
At the chosen snapshot, the star has size
\ssize$\simeq 0.0116r_c$ and the theoretical disc radius is $R_k\simeq
0.1r_c$. Note that the lopsided disc is contained within
  $R_k$. We do not expect a sharp cut-off in Fourier amplitudes beyond
  $R_k$ because the non-axisymmetric disc distorts the material beyond
  $R_k$.
 
We see that $m=1$ dominates the outer $\simeq 40\%$ of the 
disc, corresponding to the lopsided density in the contour
plot. The lopsided disc behaves like a binary companion to
the star, causing the star to orbit about their common centre of
mass. We do not expect, nor find, this motion to alter stellar
spin. The fact that $m=1$ is dominant towards the outer disc edge is
favourable for inducing orbital motion because it acts like a lever
arm.

Close to the star, $m=2$ dominates, but the contour
plot shows spiral arms that are thin and much less prominent than
those identified during initial collapse 
(Fig. \ref{case1HR_overview_snapshots}). The two arms also have
different densities (reducing the amplitude of $m=2$ symmetry). 
Well beyond the disc edge $R_k$, $m=2$ again dominates, but
this low density core material here is not expected to have
significant impact on the stellar spin because 
the total mass in this region is small,
and because the star appears more axisymmetric as
the $m=2$ component of its potential decays with distance.

The limited spin evolution is due to ineffective gravitational
coupling between the star and the disc.  
For a disc with $m$ fold symmetry, a qualitative representation
  of its instantaneous surface density is 
  \begin{align}
    \Sigma\sim
    \mathcal{A}\sin{m\phi} + \mathcal{B}\cos{m\phi},
  \end{align}
  and for a star whose potential is dominated by $k$ fold
  symmetry, we can write
  \begin{align}
    \Phi_*\sim
    \mathcal{A}_*\sin{k\phi} + \mathcal{B}_*\cos{k\phi},
  \end{align}
  where $\mathcal{A},\,\mathcal{B},\,\mathcal{A}_*,\,\mathcal{B}_*$ are
  real functions of radius.  The instantaneous torque  
  per unit area exerted on the disc by the star is 
  $\mathcal{T}=-\Sigma\partial_\phi\Phi_*$. Hence, for a disc dominated
  by the $m=1$ mode and a stellar potential with $k=2$ symmetry
  (Fig. \ref{case1HR_disc}), we expect $\langle \mathcal{T}\rangle_\phi
  \sim0$.

In other words, there is very little torque if the two
components do not share the same symmetry ($m\ne k$). Physically
  this is because there is no resonance, so for half the azimuth the
  torque is opposite sign to the other half, and averages to zero. 
 We expect a much reduced self-gravity angular momentum flux across the 
stellar surface in such situations.

\begin{figure}
  \centering
  \includegraphics[width=\linewidth]{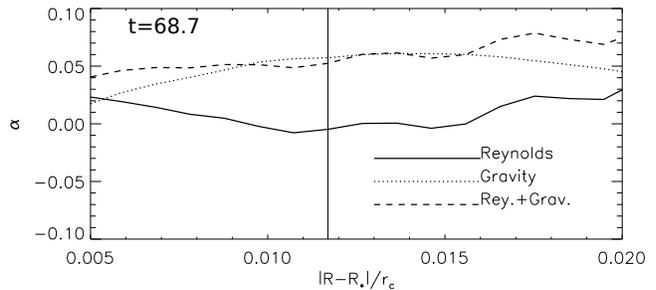}
  \caption{Case 1: non-dimensionalised, azimuthally averaged angular 
    momentum flux towards the end of the simulation. The vertical line
    is \ssize. The advective flux is $\sim -0.5$ and beyond the scale
    of this plot.  
    \label{case1HR_alpha}}
\end{figure}

Fig. \ref{case1HR_alpha} shows azimuthally averaged angular momentum
fluxes for the chosen snapshot. The self-gravity flux is
$\alpha_G\simeq 0.05$ near \ssize, and is an order of magnitude
smaller than at $t=9.25$ (Fig. \ref{case1HR_stardisc}), indicating
ineffective gravitational spin-down. This is consistent with the
limited spin evolution observed during this time, and the fact that
the star is much more axisymmetric than its bar-like shape at $t\la
11$, thus more difficult to spin down.      

We find that limited spin evolution is also correlated to significant 
motion of the star, induced by the $m=1$ mode. We now examine a system
with a smaller disc-to-star mass ratio, and consequently less orbital
motion of the star.

%%%%%%%%%%%%%%% DO NOT DELETE%%%%%%%%%%%%%%%%%%
%radius is     0.011697533
%alpha_g       1.1218635
%alpha_t    0.0074420443 (turbulent)
%alpha_t      -1.3845025 (advection)

\subsection{Case 2}\label{case2} 
This case has $m=2$ as the dominant non-axisymmetric mode  
throughout the simulation. Case 2 has parameters $\xi = 2.74$
($A=2.8$), $h=0.1$ and $q=0.01$. A lower $\xi$ corresponds to a
smaller core and disc mass compared to Case 1. Increasing $q$
corresponds to a larger star initially. Again, we wait until the star
is well-resolved at the finest grid level before making
measurements. The star size \ssize is typically resolved by at least
30 cells.

\subsubsection{Star-disc evolution}
\begin{figure}
  \centering
  \includegraphics[width=\linewidth]{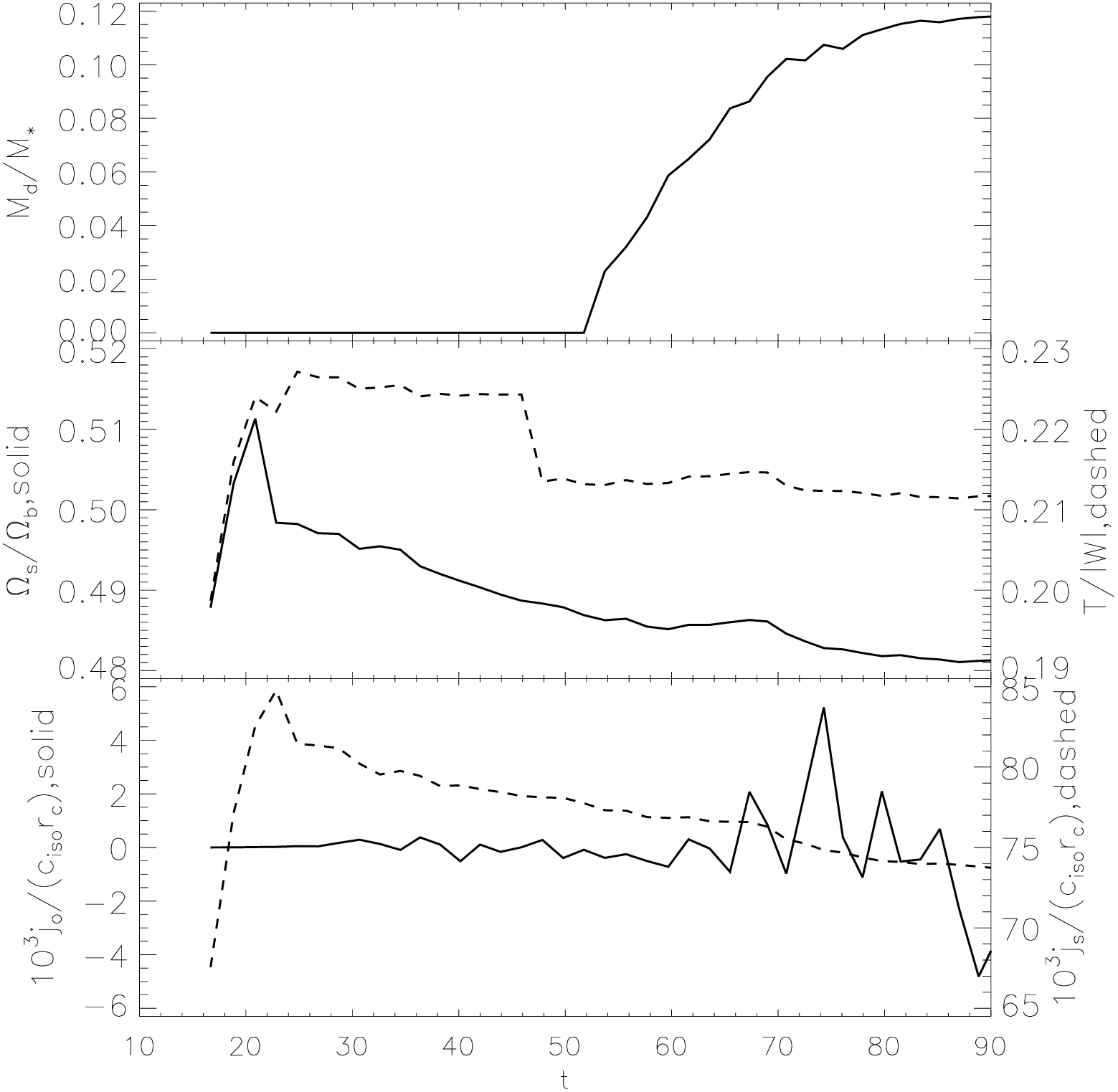}
  \caption{Case 2: evolution of the disc-to-star mass ratio
    (top), the stellar spin and kinetic-to-potential energy ratio (middle)
    and the angular momenta (bottom). Note the different scales used on
    the left and right vertical axes. 
    \label{case2HR_stardisc}}
\end{figure}

Fig. \ref{case2HR_stardisc} shows the evolution of disc mass and 
stellar properties. The phase $t\la 23$  is similar to $t\la 11$ of   
Case 1. There is an initial rapid increase in $\spin/\brk$, up to 
$\spin\simeq 0.51\brk$ followed by rapid spin-down, abruptly stopping 
at $\spin\simeq 0.498\brk$. However, unlike Case 1, for $t>23$ there
is a noticeable, almost monotonic, decrease in $\spin/\brk$. Notice the
plateau in spin frequency around $t\simeq 65$, which coincides with
increase in orbital motion (see below)  when the disc becomes
somewhat massive ($M_d\sim 0.08M_*$). Like Case 1 though, $T/|W|$
remains fairly constant. The final  spin frequency $\spin\simeq
0.482\brk$ is not much lower than the maximum value attained during
initial collapse, but the decreasing $\jspin$ suggests it may further
spin-down if the simulation were continued.

The evolution of  the spin angular momentum  is also much smoother compared to
Case 1. Unlike the previous case, for $t\ga 23$, $\jspin$ decreases
monotonically, reaching a value $\sim 10\%$ lower than the maximum at
$t\simeq23$. Here $|\jorb|$ remains at least an order of magnitude
smaller than $\jspin$, with $|\jorb|$ only growing when the disc
develops (the latter was also seen for Case 1).  

At late times, the difference in stellar spin
evolution between the present case and Case 1 is likely due to very
different disc properties. In Case 2, $M_d/M_*\la 0.12$, smaller than
Case 1 by a factor of $\sim 4$. A disc with large $M_*/M_d$ is
expected to be stable against the growth of $m=1$ perturbations
\citep{heemskerk92}, because there is insufficient disc mass to move
the star.   

\begin{figure}
  \centering
 \includegraphics[width=\linewidth]{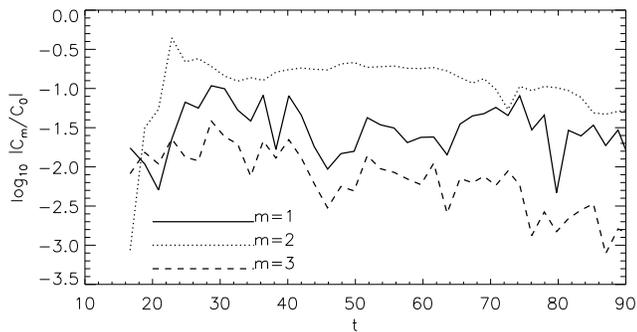} 
  \caption{Evolution of non-axisymmetric modes in the surface density
    of material external to the star. Fourier amplitudes were
    integrated over a cylindrical distance  $[0.04,0.25]r_c$ away from
    the star, then normalised by the  axisymmetric amplitude.   
    \label{case2_discmodes}}
\end{figure}

Fig. \ref{case2_discmodes} shows the evolution of non-axisymmetric
modes in surface density and is very different to Case 1. Here,
the $m=2$ mode remains dominant throughout the simulation. This is
consistent with the idea that the $m=2$ mode provides the 
spin-down torque while the $m=1$ mode produces orbital motion.

\begin{figure} 
    \centering 
    \includegraphics[width=\linewidth]{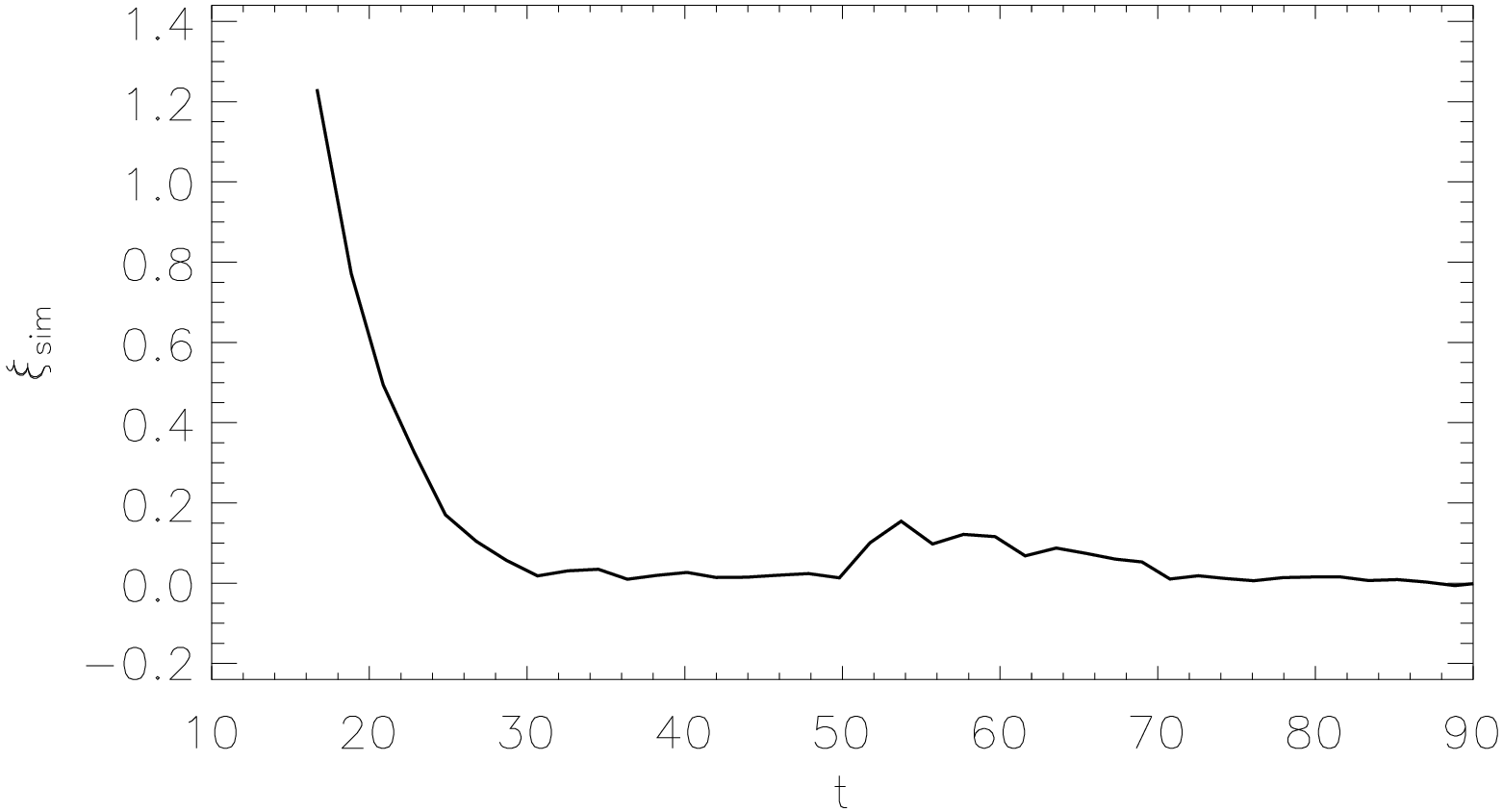}
    \caption{Case 2: the dimensionless mass accretion rate. Note that our
      measurements begin after the central object is well
      resolved. The initial phase observed in Case 1, where
      $\xi_\mathrm{sim}$ is roughly constant, is not present
      because for this simulation, the central object is not yet well
      resolved during these very early times. 
      \label{case2HR_mass_acc}}
  \end{figure}

  Fig. \ref{case2HR_mass_acc} shows the measured mass accretion rate. 
  $\xi_\mathrm{sim}$ is smaller than Case 1 because of the lower mass 
  core. As before, there is a drop in accretion associated with
  self-limited spin. At late times, $\xi_\mathrm{sim}$ is
  roughly constant but is much smaller than $\xi$. This could be
  because we have a disc which is receiving angular momentum from the
  spin-down of the central star. This can be thought of as an accretion
  disc with a positive torque applied at the inner boundary
  \citep{yuan85}. This effect is absent in \cite{kratter10}, where 
  $\xi$ remains close to its initialisation value, because in those simulations 
  the central object was represented by a sink particle that was not capable 
  of torquing the disc.

%   In Fig. \ref{case2HR_mass_flux} we show the spatial dependence of
%   mass flux at three times. At $t=18.9$, the central object has
%   average radius $\sim 0.03r_c$. The positive flux gradient just
%   outside $0.03r_c$ in 
%   Fig. \ref{case2HR_mass_flux} then indicate the central object to be
%   accreting material from just outside its surface. Hence,  this
%   snapshot coincides with spin-up (Fig. \ref{case2HR_stardisc}). 

%   In Fig. \ref{case2HR_mass_flux}, the curve for $t=26.8$, correspond
%   to when the central object has spun-down after reaching maximum
%   spin (Fig. \ref{case2HR_stardisc}). Here, mass fluxes are
%   lowered considerably, consistent with increased difficulty for
%   external material to join the central object if the former is
%   receiving angular momentum from the latter. At later times,
%   $t=59.7$, the mass flux is negligible compared to the initialisation
%   parameter $\xi$, probably because we now have a disc supported by a
%   central mass, which is receiving angular momentum from the spin-down
%   of the central object. As pointed out by \cite{yuan85}, this
%   can be thought of an accretion disc with a positive torque
%   applied at the inner boundary.   

%   \begin{figure} 
%     \centering 
%     \includegraphics[width=\linewidth]{figs/case2d_mdot_flux.ps}
%     \caption{Case 2: non-dimensionalised mass accretion rate. 
%       \label{case2HR_mass_flux}}
%   \end{figure}

%-------------------------------------------------------

\subsubsection{Disc structure and angular momentum flux}
\begin{figure}
  \includegraphics[width=\linewidth]{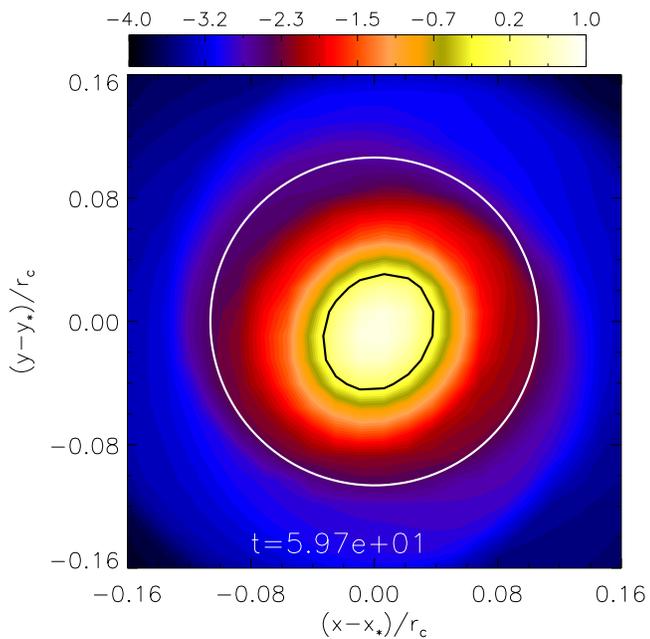} 
  \caption{Case 2: snapshot of $\log{(\rho/\rho_*)}$ in the
      star's equatorial plane. The inner black line indicates
      $\rho=\rho_*$ and the outer white line indicates $R_k$.
    \label{case2HR_structure}}
\end{figure}

\begin{figure}
  \includegraphics[width=\linewidth]{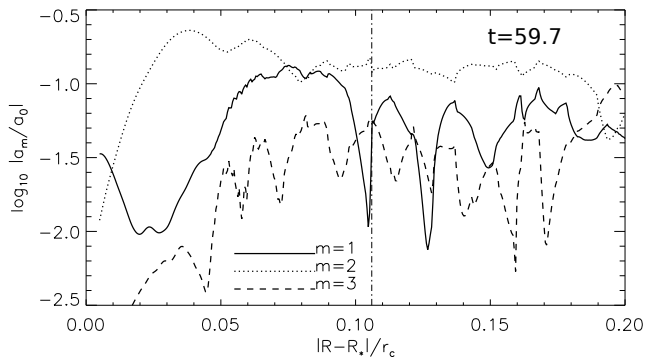}
  \caption{Case 2: Fourier 
    amplitudes of non-axisymmetric modes in the surface density 
    corresponding to Fig. \ref{case2HR_structure}. The 
    predicted disc edge is marked by the vertical line (right). The 
    star has an average radius \ssize$\simeq0.036r_c$. 
    \label{case2HR_structure_FFT}}
\end{figure}

We examine Case 2 at $t=59.7$, during its long-term spin
down. Fig. \ref{case2HR_structure} and
Fig. \ref{case2HR_structure_FFT} show the density field in real and
Fourier space, respectively. The average radius of the star is
\ssize$\simeq 0.036r_c$ and the theoretical disc edge  is $R_k\simeq
0.106r_c$. The contour plot shows that the bulk of the disc is within
  $R_k$ of the star, but it distorts the region beyond it.

The contour plot clearly shows the star-disc system has $m=2$
symmetry. Note that the relative Fourier amplitudes are quite
different from Case 1, as the $m=1$ mode is never dominant. Instead
the $m=2$ mode dominates most of the disc, and even the region beyond
$R_k$.

This explains why Case 2 experiences spin-down whereas Case 1 does 
not. In both cases the star has $m=2$ symmetry, but only in Case 2
does the disc also have a significant $m=2$ amplitude. Furthermore, the 
$m=1$ amplitude in Case 2 is smaller than that of Case 1, consistent
with limited orbital motion. The lack of a strong $m=1$ mode appears
favourable for spin-down via gravitational torques.  

\begin{figure}
%  \centering
  \includegraphics[width=\linewidth,clip=true, trim=0cm 1.5cm 0cm
  0cm]{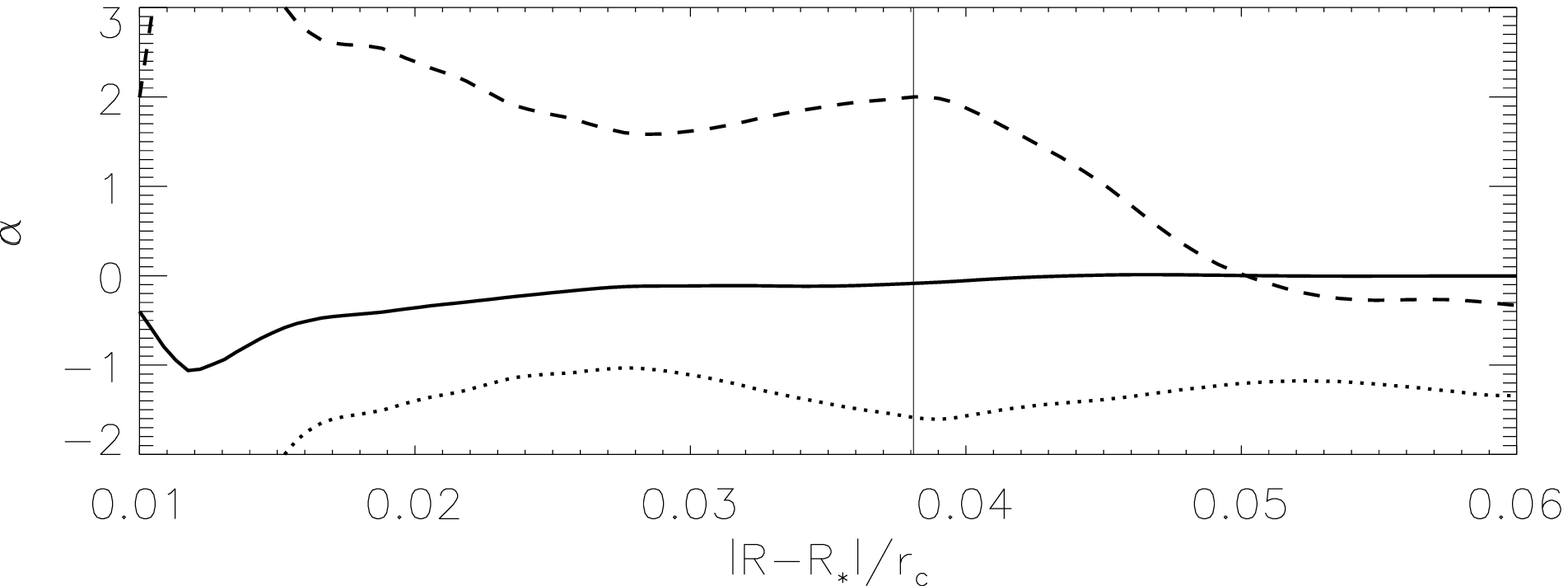}
 \includegraphics[width=\linewidth,clip=true,trim=0cm -0.5cm 0cm
   0.0cm]{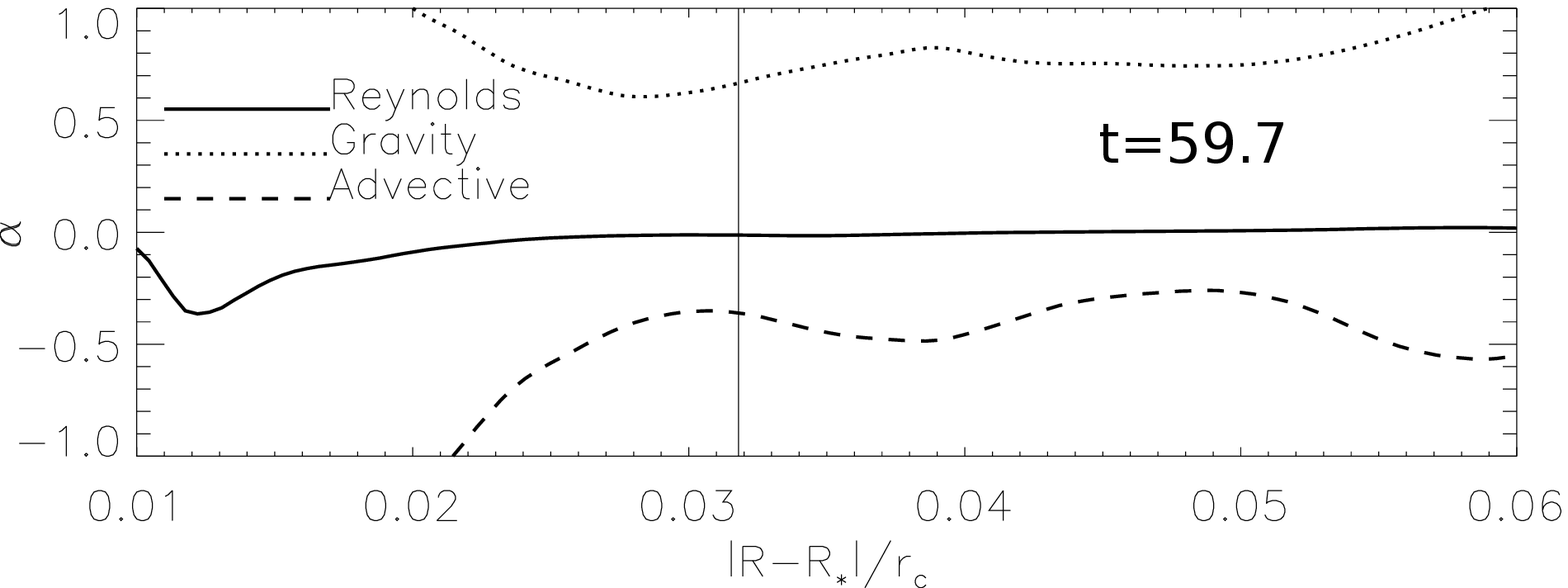} 
  \caption{Case 2: non-dimensionalised angular momentum fluxes
    during the slow spin-down phase. Top: fluxes along the
    $\phi=\pi/4$ azimuth (approximately the long axis of the star.  
    Bottom: fluxes along the $\phi=3\pi/4$ (approximately the short
    axis of the star). In each, the vertical line
    indicates the surface of the star along that
    direction. Note the different scales used for the two panels. 
    \label{case2HR_alpha}}
\end{figure}

Fig. \ref{case2HR_alpha} shows angular momentum fluxes along the long 
and short axis of the star in its equatorial plane. In both
cases, Reynolds stresses are negligible compared to gravity or
advection. The gravity flux is large, 
with $\alpha_G=O(1)$. Note the sign of the fluxes depends on the
azimuthal direction. Averaging the fluxes over the non-axisymmetric
radius where $\rho=\rho_*$, we found $\alpha_G\sim 0.3,\,
\alpha_R=O(10^{-3})$ and the large-scale advective flux $\alpha_A\sim
-0.1$. The numerical viscosity is $\alpha_N =  O(10^{-3})$. The total 
non-dimensional flux is $\sim 0.2$, consistent with the observation of
spin angular momentum loss.

%%%%%%%%%%%%%%%DO NOT DELETE%%%%%%%%%%%%%%%%%%%
%radius is     0.036398627
%alpha_g      0.34455154
%alpha_t    0.0028865219 (turbulent)
%alpha_t     -0.10220106 (advection)

\begin{figure}
  \centering
  \includegraphics[width=\linewidth]{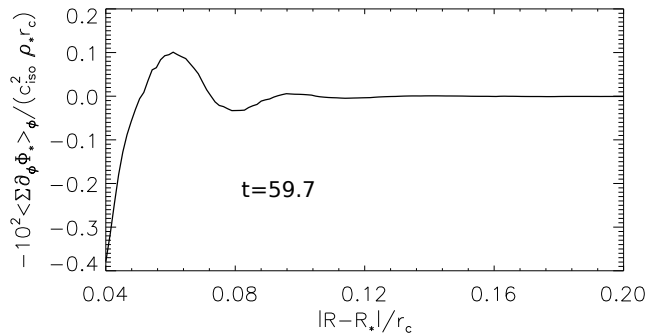} 
  \caption{Case 2: azimuthally averaged torque per unit
    area exerted on the disc by the star, during the slow spin-down
    phase.  
    \label{case2HR_torque1d}}
\end{figure}

We computed the potential $\Phi_*$ to find the torque exerted by
  the star on the surrounding disc, shown in
Fig. \ref{case2HR_torque1d}. The torque becomes negative and large in 
amplitude towards $0.04r_c$, or approximately the stellar surface.
This large negative torque might lead to the impression that the 
disc is spinning up the star, contradicting the observed spin-down,
but care must be taken when interpreting azimuthally averaged,
one-dimensional profiles such as Fig. \ref{case2HR_torque1d}. Since
the star-disc interface is non-axisymmetric, there is no single value
of radius to represent the star-disc interface. However, the
azimuthally-averaged torque is positive around $0.06r_c$ from the
star, which is certainly in the disc region (see
Fig. \ref{case2HR_structure} where this region has typical densities
much lower than the transition density). This implies a net loss of
spin angular momentum from the star to this disc region.

\section{Discussion}\label{caveats}
In our simulations, we observe an upper limit to stellar spin
  that is below break-up speed. This limitation occurs during the
  initial spin-up of the star. This relies on the 
  ability of the central object to deform into a non-axisymmetric
  shape and lose angular momentum to its surroundings by exerting
  gravitational torques on the external medium.  

  Consider a rotating, axisymmetric self-gravitating fluid body of
  mass $M$, angular momentum $J$ and radius $a$ in the plane
  perpendicular to the spin axis. We define the dimensionless
  angular momentum as  
  \begin{align}\label{critical_ratio}
    \mathcal{R} \equiv \frac{J}{\sqrt{GM^3 a}}.
  \end{align}
  Now, specialise to a Maclaurin spheroid with angular
  momentum $J=(2/5)a^2\omega M$, where $\omega$ is the uniform spin
  angular frequency. The dimensionless parameter $\mathcal{R}$ for the
  Maclaurin spheroid is  
  \begin{align}\label{critical_ratio2}
    \mathcal{R} =
    \frac{2}{5}\left[\omega\left/\sqrt{\frac{GM}{a^3}}\right.\right]\equiv \mathcal{R}_\mathrm{Mac}.   
  \end{align}
  If $\mathcal{R}_\mathrm{Mac}\ga 0.278$, triaxial equilibrium
  with uniform rotation is possible \citep{yuan85}. The quantity in
  square brackets is equivalent to our spin parameter
  $\spin/\brk$. If we increase $\mathcal{R}_\mathrm{Mac}$
  from a small value, the Maclaurin spheroid will exceed the critical
  value above \emph{before} it reaches break-up speed (at which point
  $\mathcal{R}_\mathrm{Mac}=0.4$). That is, it may become
  non-axisymmetric before flying apart. Obviously, our central object
  is not a Maclaurin spheroid, but inserting the observed maximum spin
  for Case 1, $\spin/\brk\sim 0.55$, yields $\mathcal{R}_\mathrm{Mac}= 
  0.22$, which is not too different from 0.278.

  This analysis makes an important point which, together with our
  our simulations, suggests that the maximum stellar spin may be
  determined largely by the critical value of $\mathcal{R}$ (or
  $\spin/\brk$), beyond which instabilities in the rotating object
  produce non-axisymmetric deformations. As soon as this occurs, the
  object may torque the surrounding disc and lose angular momentum to
  it. The transition from Maclaurin spheroids to Jacobi ellipsoids
  \citep{chandrasekhar69} at $\mathcal{R} = 0.278$ is one example, but in
  general the critical value will depend on the properties of the
  central object, such as its equation of state \citep{ostriker73}.  
  We may think of objects with lower critical values of $
  \mathcal{R}$ as being easier to deform, and will likely
  be limited to smaller maximum spin rates. For example, our
  compressible star with smaller $\mathcal{R}_\mathrm{Mac}$
  is easier to deform than the incompressible Maclaurin spheroid.

  In our simulations we treat the star as a simple $n=3/2$ polytrope,
  but this is obviously an oversimplification of true protostellar
  structure, which will in general depend on its mass and
  evolutionary state. Even if we were to include a more realistic
  treatment of the EOS for stellar material, our limited resolution
  would make it impossible to study the star's deformation in anything
  but a very schematic manner. Such a study is beyond the scope of this
  paper, but due to this limitation we should be cautious about putting
  too much weight on the exact numerical values we compute for the
  gravitational spin-down torque. Our general conclusions that the
  initial spin-up is self-limited to well below breakup, and that a
  prominent $m=2$ mode in the surrounding disc is capable of providing a
  further, long-term spin-down torque, should be robust.
  In
  particular, it is important to note that the argument we have given
  above is completely dimensionless, and does not depend in any way on
  the true physical sizes of protostars or their discs.

  A related issue is the degree of deformation needed for effective
  angular momentum exchange between the star and its surroundings. In
  our simulation, the density changes smoothly across the star-disc
  interface, whereas in reality radiation from both the accretion shock
  and the stellar surface reduce the entropy of stellar material well
  below that of disc material. This causes stellar matter to be far
  denser than matter in the disc immediately outside the star, an effect that
  we miss because our simulations are non-radiative. 
  A larger density
  contrast would make the star's gravitational field stronger, and would
  therefore increase the effectiveness of gravitational angular momentum
  transport relative to that in our simulations. As a result, real stars
  probably spin down more effectively than we have found, and require
  less non-axisymmetry to achieve the same torque
  \cite[c.f.][]{yuan85}. Thus a more realistic treatment of stellar
  structure may actually strengthen the effect we have
  identified.

\subsection{Numerical issues}  

The issue of numerical viscosity was not discussed in detail, 
which could be important for numerical studies of
rotating fluids in a Cartesian box. Angular momentum conservation
can be violated near the centre of the rotating fluid.  However, we
have re-calculated spin angular momenta excluding the innermost few 
cells and found negligible effect on their evolution. In addition,
lower resolution runs were performed which yield similar angular
momenta evolution. We are thus confident that the observed correlation between spin
evolution and evolution of disc modes are physical results.

We have focused on only two simulation runs. Although
insufficient to predict stellar spin evolution as a function of the problem
parameters, our results provide an useful guide to future numerical
studies. We expect discs that develop prominent global $m=2$
spirals to be most effective at spinning down the star
through gravitational torques, because the star will usually have the
same symmetry.

\subsection{Additional caveats}

It should be noted that the physics in our problem setup is highly
simplified. The dynamical range covered in our simulations does
not correspond to an entire, realistic star-disc system. This
discrepancy could potentially limit the applicability of our results
to realistic star-disc systems. 

One issue is that if the star-on-disc torque is most significant close
to the stellar surface, then such torques will be ineffective
if there exists a significant gap between the star and the inner disc
edge. A gap may be caused by magnetic fields for low mass
stars. However, using the model of \cite{matt05}, one finds that for
massive stars the truncation radius is actually inside the star,
implying no gap \citep{rosen11}. Gravitational torques may then play a
role. In addition, if gravitational torques are a result of high
protostar spin rate, which enables triaxiality, then material must have
fallen onto the star, adding to its angular momentum in the first
place. The smooth star-disc boundary in our simulations is then
self-consistent with limiting spin via gravitational torques in the
early stages of collapse. 

Finally, although we have demonstrated the existence of star-disc
torques, we caution that the exact magnitudes of the torques will
likely depend on the structure of the disc, which can in turn be
modified by magnetic fields and radiation. We have not included these
effects, and in their presence the true spins of stars undergoing
gravitational torque-regulated spin-up as they form will likely be
quantitatively different than what we have found, but our general
conclusions should hold.

\section{Conclusions}\label{conclusions}
We performed hydrodynamic simulations of the collapse of an
  isothermal cloud leading to the formation of a disc with a dense
  central object. We focused on the evolution of the central
  object's spin and its relation to disc properties.  

We find that the initial spin-up of the central object does not
exceed $\sim 50\%$ of its break-up speed because increasing spin also
increases its deformation into a bar-like object (possibly
through an instability), on which the external material exerts a
negative torque. Spin-down occurs over a period that is short
compared to star formation timescales.

We also find that when the surrounding disc develops a dominant 
$m=1$ disturbance, the central object's spin evolution is
inhibited, and large orbital motions are induced by the disc. Our
experiments suggest, perhaps counter-intuitively,  that massive discs
are less able to provide long term  gravitational spin-down torques on
a  star because such discs are prone to developing $m=1$ lopsided 
modes.

Although spin-down
was observed over the simulation timescale for systems
with $m=2$ symmetry, the spin-down rate is small
compared to that during the earliest phase of collapse.  
The precise value of this upper limit
may depend on the internal structure of the central object  
in ways that we have not explored. In general we expect that
structures that are easier to deform, in the sense that a smaller
  spin-to-break-up frequency ratio is needed to allow non-axisymmetry 
  , will yield smaller maximum spin rates than structures
  that require larger spin-to-break-up frequencies to become
  non-axisymmetric. 

An upper limit to spin rate
has important implications for stellar
evolution. Strong rotational mixing may occur for large spin rates and
allow the star to bypass the red giant phase
\citep{woosley06,yoon05}. The critical spin rate found by
\citeauthor{woosley06} is $\simeq 40\%$ of break-up. As another
example, \cite{ekstrom08} found that high rotation ($\simeq
40\%$---70\% break-up) can increase metal production in initially
metal-free stars. However, if
gravitational torques limit stellar spin to $\simeq 50\%$ of
break-up, strong rotational mixing or increased metal production may
be difficult to achieve because once the star reaches the main sequence,
stellar winds provide further spin-down. A maximum stellar spin set by
gravitational torques can be used to constrain the parameter space for
stellar evolution calculations.  
\\\\

This project was initiated during the ISIMA 2010 summer program, funded by the
NSF CAREER grant 0847477, the France-Berkeley fund, the Institute for
Geophysics and Planetary Physics and the Center for Origin, Dynamics and
Evolution of Planets. We thank them for their support. MKL
acknowledges support from the Isaac Newton Trust and St. John's
College, Cambridge. MRK acknowledges support from: an Alfred P. Sloan
Fellowship; NSF grants AST-0807739 and CAREER-0955300; and NASA
through Astrophysics Theory and  Fundamental Physics grant NNX09AK31G
and a Spitzer Space Telescope Theoretical Research Program grant. KMK is supported by
an Institute for Theory and Computation Postdoctoral Fellowship at Harvard College Observatory.

%\bibliographystyle{mn2e}
%\bibliography{ref}	

%\begin{thebibliography}
%\end{thebibliography}

\end{document}